\def\Journal#1#2#3#4{{#1} {\bf #2}, #3 (#4)}

\def\PLB{{ Phys. \mbox{Lett.}}  B}
\def\PRL{ Phys. Rev. \mbox{Lett.}}
\def\PRD{{ Phys. Rev.} D}

\newcommand{\etal}{{\it et al.}~}
\newcommand{\reac}{$e p\to e n \pi^+ ~$}
\newcommand{\reacX}{$e p\to e \pi^+ X ~$}	
\newcommand{\reacpimX}{$e p\to e \pi^- X ~$}

\documentclass[twoside]{dis04}

\begin{document}

\title{Exclusive $\pi^+$ production at HERMES
}

\author{\underline{CYNTHIA HADJIDAKIS}$^\mathrm{a}$, DELIA HASCH$^\mathrm{a}$ and 
ERIC THOMAS$^\mathrm{b}$ \\ (on behalf of the HERMES Collaboration)}

\address{$^\mathrm{a}$INFN, Laboratori Nazionali di Frascati, 00044 Frascati, Italy\\
$^\mathrm{b}$Laboratory for High Energy Physics, University of Bern, 3012 Bern, Switzerland\\
E-mail: cynthia@lnf.infn.it, hasch@lnf.infn.it, eric.thomas@cern.ch}

\maketitle

\abstracts{
Hard exclusive production in deep inelastic lepton scattering provides 
access to the unknown Generalized Parton Distributions (GPDs)
of the nucleon.
At HERMES, different observables for hard exclusive 
$\pi^+$ production have been measured with a 27.6 GeV 
positron beam on an internal hydrogen gas target. 
First preliminary results for the unpolarized \reac total cross section 
for $1.5<Q^2<10.5$ GeV$^2$ and for $0.02<x<0.8$ are presented and 
compared to GPD calculations. 
The final result 
for the single-spin asymmetry using a longitudinal polarized 
target is also reported.
}

\section{Introduction}	
The interest in the hard exclusive electroproduction of mesons 
has grown since a QCD factorization 
theorem was proved in the case of longitudinal photons~\cite{Collins}. 
It was shown that at large $Q^2$ and low $t$, the amplitude for such reactions can be factorized into a hard lepton scattering part, which can be calculated in perturbative QCD, and two soft parts which parametrize the produced meson by a distribution amplitude and the target nucleon by four Generalized Parton Distributions (GPDs)~\cite{GPD}. 
The quantum numbers of the produced meson select different GPDs. While exclusive vector meson production is only sensitive to unpolarized GPDs ($H$ and $E$), pseudoscalar meson production is sensitive to polarized GPDs ($\tilde H$ and $\tilde E$) without the need for a polarized target or beam. Moreover, in the case of $\pi^+$ production, 
the pseudoscalar contribution $\tilde E$ is dominated at low $t$ by the 
pion-pole exchange, and therefore $\tilde E$ is related to the pion form 
factor. Different observables, like cross section and single-spin asymmetry are able to select different combinations of GPDs.

In this paper, the first preliminary measurement of the total cross section 
for the exclusive $\pi^+$ production is reported.
The final result for the single-spin asymmetry on a longitudinally polarized 
target is also presented.
The data were collected using an internal hydrogen gas target in the 
27.6 GeV HERA positron storage ring at DESY. 

\section{Total cross section measurement}

The scattered positron and the produced hadron were detected by the HERMES 
spectrometer~\cite{HERMES} which features excellent lepton-hadron
separation and good pion identification for momenta between 2.5 and 15 GeV.
The kinematic requirements imposed on the scattered positrons were: $Q^2>1$ 
GeV$^2$, $0.02<x<0.8$ and an invariant mass squared of the initial 
photon-nucleon system $W>2$ GeV.
The recoiling neutron was not detected, and exclusive production 
of mesons was selected by requiring that the missing mass square ($M_X^2$) 
of the reaction \reacX correspond to the nucleon mass square. 
Due to the limited experimental resolution, the exclusive $\pi^+$ reaction 
cannot be separated from the neighboring channels (defined as 
non-exclusive) which can be smeared into the exclusive region. Therefore 
the process \reacpimX was used to subtract non-exclusive 
channels~\cite{SSA_paper}. Indeed, 
exclusive production of $\pi^-$ on a hydrogen target with a recoiling nucleon in the 
final state is forbidden due to charge conservation. 
The background correction procedure was also tested using a small data sample 
taken with a 12 GeV beam energy where the relative experimental resolution 
was higher.
The exclusive peak resulting from the background correction 
was centered at the nucleon mass for both 12 and 27.6 GeV data samples,
confirming the reliability of the procedure.

For the total cross section measurement, data from 1996 to 2000 on 
unpolarized and polarized hydrogen targets were used. 
Approximately 3.5 k exclusive $\pi^+$ were selected. The $Q^2$ dependence 
of the cross section has been determined for three different bins in $x$ 
following:
\begin{eqnarray}
\nonumber
\sigma^{\gamma^* p\to n \pi^+ }(x,Q^2)&=&\frac{N^{excl}_{\pi^+}(x,Q^2)}
{L \Delta x \Delta Q^2 \Gamma_V(\langle x \rangle,\langle Q^2\rangle) \kappa(x,Q^2)}
\end{eqnarray}
\noindent
where $N^{excl}_{\pi^+}$ is the number of $\pi^+$ corrected for background,
 $\Gamma_V$ is the virtual photon flux factor and $\kappa$ is the 
probability to detect the scattered positron and the produced 
$\pi^+$ with the HERMES 
spectrometer. These quantities were determined for each $Q^2$ 
and $x$ bin. 
$L$ is the integrated luminosity and the symbol $\Delta$ is related to the 
size of the bin.
\begin{figure}[t]
\begin{center}
\includegraphics[width=0.75 \textwidth]{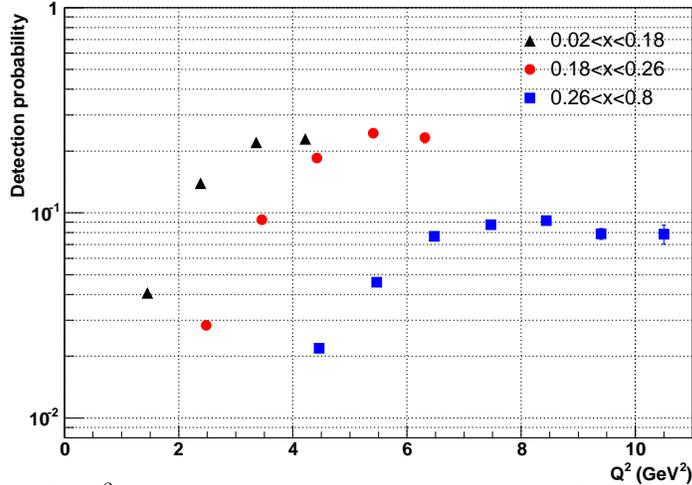}
\vspace*{-0.5cm}
\caption[]{
$Q^2$ dependence of the detection probability for different $x$ ranges.}
\label{fig1}
\end{center}
\end{figure} 
The detection probability has been determined using two exclusive Monte Carlos 
based on two different GPD parametrizations~\cite{Piller}~\cite{VGG}.
The $t$ distribution, 
which is the most relevant observable for exclusive production, is 
well described by both Monte Carlos.
The $Q^2$ and $x$ distributions are 
better described by~\cite{VGG}. For this reason, the latter 
parametrization was used to determine the detection probability and 
the parametrization by~\cite{Piller} 
served to estimate the systematic error. 
The calculated detection probability is shown in 
Fig.~\ref{fig1}. It varies between 0.02 and 0.25 according 
to the $Q^2$ and $x$ bin with a relative systematic error 
evaluated to be lower than $15\%$. 
The total systematic error on the total cross section 
is dominated by the uncertainty on the 
background subtraction, 
and by the uncertainty on the efficiency correction due to different 
detector resolutions between 1996 and 2000.
\begin{figure}[t]
\begin{center}
\includegraphics[width=0.8 \textwidth]{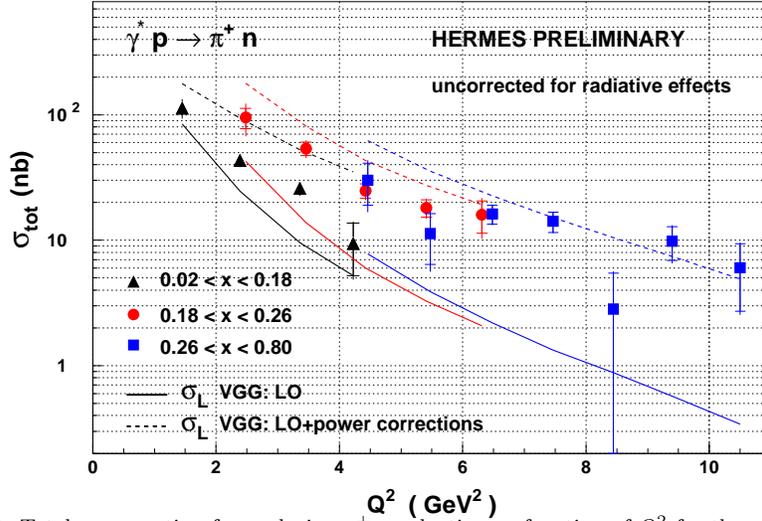}
\vspace*{-0.5cm}
\caption[]{Total cross section for exclusive $\pi^+$ production 
as function of $Q^2$ for three different $x$ ranges and integrated over $t$. 
The inner error bars 
represent the statistical uncertainty and the outer error bars the quadratic 
sum of statistical and systematic uncertainty. The curves represent 
calculations based on a GPD-model~\cite{VGG}.}
\label{fig2}
\end{center}
\end{figure}
Fig.~\ref{fig2} shows the $Q^2$ dependence of the total cross section 
for three different $x$ ranges.
These preliminary data have not yet been corrected for radiative effects. This correction 
has been roughly estimated to be as large as 20$\%$ showing almost 
no dependence on $x$ or $Q^2$.
The data have been compared to calculations for the longitudinal part of the 
cross section computed by GPD-model~\cite{VGG}. 
The total cross section can be written as 
$\sigma = \sigma_T + \epsilon \sigma_L$ where $\sigma_T$ ($\sigma_L$) 
is the transverse (longitudinal) virtual photons contribution and $\epsilon$ 
is the virtual photon polarization parameter.  
At HERMES the separation of the transverse and longitudinal component of 
the cross section is not feasible.
However, since the transverse contribution is predicted to be suppressed 
by a power of $1/Q^2$ with respect to the longitudinal one~\cite{Collins}
and since $\epsilon$ ranges from 0.8 to 0.95, the data at larger $Q^2$
are expected to be dominated by the longitudinal part.
The full lines in Fig.~\ref{fig2} 
show the leading-order calculation computed for the mean $x$ and $Q^2$ 
corresponding to the data and integrated over $t$.
The dashed lines include 
power corrections due to intrinsic transverse momenta 
of the partons in the nucleon and due to soft overlap type contributions.  
The $Q^2$ dependence is in general agreement with the theoretical 
expectation. While the leading-order calculations underestimate the data, 
the evaluation of the power corrections appears too large.

\section{Single-Spin Asymmetry measurement}

It has been predicted \cite{fpsv} that for 
the exclusive production of $\pi^+$ mesons 
from a transversely polarized target by longitudinal
virtual photons,
the interference between the 
pseudoscalar ($\tilde{E}$) and pseudovector ($\tilde{H}$) amplitudes leads
to a large target single-spin asymmetry.
While the unpolarized cross section depends on a quadratic 
combination of the two polarized GPDs, a linear dependence 
of these distributions appears in the single-spin asymmetry. 
Polarization is therefore needed in order to disentangle the 
different contributions.
Moreover, the scaling region of the asymmetry (where
corrections proportional to powers of $1/Q$ are small) is expected to be 
reached at lower $Q^2$ than for the absolute 
cross section.
It has also been shown that corrections that are next-to-leading order 
(NLO) in $\alpha_s$ cancel in the transverse
asymmetry \cite{fpps}.

Since 2002, data have been collected at HERMES using a transversely polarized 
hydrogen gas 
target~\cite{trans}. This running period will continue till 2005 
and the expected 5 millions deep inelastic scattering events will allow one,
 with more than $10^3$ resulted exclusive $\pi^+$ events, 
to measure the single-spin asymmetry 
on a transversely polarized target.

HERMES has already measured the single-spin asymmetry 
in the exclusive production of 
$\pi^+$ using a {\it longitudinally} polarized hydrogen target with data 
collected in 1997~\cite{SSA_paper}. 
The cross section asymmetry for exclusively produced $\pi^+$ is defined by
\begin{eqnarray}
\nonumber
A(\phi) = \frac{1}{|S|} \,\frac{N_e^{\uparrow}(\phi) - N_e^{\downarrow}(\phi)}{N_e^{\uparrow}(\phi) + N_e^{\downarrow}(\phi)}.
\label{csa_e}
\end{eqnarray}
\noindent
where $N_e$ represents the yield of exclusive $\pi^+$, the superscript
$\uparrow(\downarrow)$ denotes a target polarization direction 
anti-parallel (parallel) to the positron beam momentum,
and $S$ is the degree of polarization of the target protons 
of $0.88 \pm 0.04$. Here $\phi$ is the azimuthal angle of the pion 
around the virtual photon momentum
relative to the lepton scattering plane.
\begin{figure}[t]
\begin{center}
  \epsfig{file=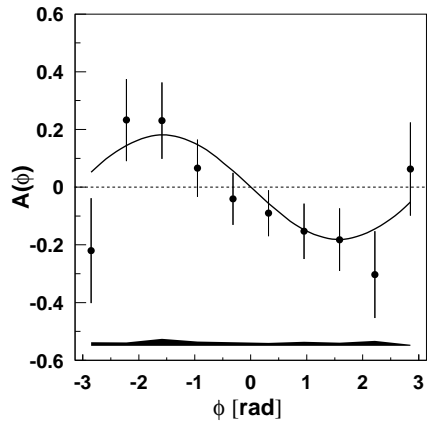, width=5.8cm, angle=0}
\vspace*{-0.5cm}
\caption{Cross section asymmetry for the reaction
$e \vec{p} \rightarrow e n \pi^+$. The curve is a fit 
to the data by $A_{\mathrm{UL}}(\phi) = A_{\mathrm{UL}}^{\sin \phi} \sin \phi$
which yields $A_{\mathrm{UL}}^{\sin \phi} = -0.18 \pm 0.05 \pm 0.02$.
}
\label{fig3}
\end{center}
\end{figure}
The cross section asymmetry integrated over $x$, $Q^2$ and $t$ is shown 
in Fig~\ref{fig3} .The average values of the kinematic variables are
$\langle x \rangle = 0.15$, $\langle Q^2 \rangle=2.2\;\mbox{GeV}^2$ and 
$\langle t \rangle = -0.46\;\mbox{GeV}^2$.
The data show a large asymmetry in the distribution 
versus azimuthal angle $\phi$, with 
a clear $\sin \phi$ dependence.
A fit to this dependence of the form
$A(\phi) = A_{\mathrm{UL}}^{\sin \phi} \cdot \sin \phi$
yields $A_{\mathrm{UL}}^{\sin \phi} = -0.18 \pm 0.05 \pm 0.02$.

In the case of electroproduction from a target polarized \textit{longitudinally}
 with respect to the lepton beam momentum, the data are dominated by 
the contribution from the longitudinal target polarization component with 
respect to the virtual photon direction, therefore 
a comparison to the predictions 
for a \textit{transversely} polarized target would require a calculation with the 
inclusion of next-to-leading twist contributions~\cite{fpps}.

\section{Conclusion and prospects of exclusive pseudoscalar measurements at HERMES}

Results have been presented for hard exclusive $\pi^+$ production 
performed with the HERMES spectrometer.
The preliminary total cross section has been measured 
as a function of $Q^2$ for 
different $x$ values and has been compared 
to calculations based on a GPD-model. 
Also, a significant negative single-spin asymmetry has been observed with a 
longitudinally polarized target. 
The running period with a transverse target polarization 
which will continue till 2005 will allow one to measure the target-spin 
azimuthal asymmetry for exclusive $\pi^+$ and to make 
a direct comparison to the GPD-model predictions.
Furthermore, with the installation of the recoil detector 
around the target in 2005, 
it will become possible to detect the recoil proton. This will allow one to 
measure neutral pseudoscalar meson exclusive reactions on hydrogen and to 
better identify the charged meson ones. 
In particular, the study of the ratio for different pseudoscalar mesons 
($\pi^+$ over $\pi^0$, $\pi^0$ over $\eta$, ...) will 
provide information on the different contributions with respect to the 
nature of the produced mesons. 

\section*{Acknowledgements}
We like to thank Nicola Bianchi for very interesting and helpful discussions.

\end{document}